# A Non-Local Model for Fracture Closure on Rough Fracture Faces and Asperities*

HanYi Wang, and Mukul. M. Sharma
Petroleum & Geosystems Engineering Department, The University of Texas at Austin, United States

**Abstract**

Natural fractures, hydraulic fractures, and acid etched fractures have some degree of fracture surface roughness. These surface asperities are largely responsible for the hydraulic conductivity of these fractures. This paper presents a model to quantify the fracture closure process that is crucial to predicting the stress dependent conductivity of fracture networks in unconventional reservoirs and estimating the minimum in-situ stress using fracture injection tests. Past studies that have investigated the fracture closure process have assumed the fracture surfaces to be two parallel plates closing in uniformly on rough surfaces and asperities. In reality hydraulically induced fractures are wider in the middle and narrower near the fracture tip. As a result, asperities on the rough fracture surfaces come into contact near the fracture tip well before they do near the middle of the fracture. The evolution of the entire fracture geometry and its impact on stress redistribution and dynamic fracture closure behavior has not yet been investigated.

In this paper, we present a method and general algorithms to model the dynamic closure behavior of a hydraulic fracture while accounting for the initial fracture shape, rough fracture surfaces and deformation of asperities in contact. Analytic solutions from linear elastic fracture mechanics for three fracture models (PKN, KGD and radial fracture geometry) are coupled with a general contact law to show that the fracture closure process is a gradual, non-local process, which occurs at the fracture edges initially and then moves progressively to the center of fracture, as the fluid pressure inside the fracture declines. Our study also reveals that the minimum in-situ stress should not be picked at the occurrence of mechanical closure as conventional practice suggests.

**Keywords:** hydraulic fracture; fracture closure; surface roughness; asperities; stress determination; surface contact

## 1. Introduction

Hydraulic fracturing has been widely used to enhance the recovery of hydrocarbons from very low permeability reservoirs, as well as prevent sand production in high permeability reservoirs (Economides and Nolte 2000). Development of theoretical models of hydraulic fracturing started several decades ago, and provided a basis for hydraulic fracturing design, optimization and diagnostics. Traditionally, hydraulic fracture models assume that the fracture can be represented as a slot between smooth, flat surfaces. This assumption is based on observations that most fractures are approximately planar on the scale of the fracture length, and analytical solutions to the laminar flow problem in such slots can be obtained. Models for hydraulic fracture propagation in rocks have been used extensively, from early 2D analytic models, such as such the Khristianovic-Geertsma-de Klerk (KGD) model (Geertsma and De Klerk 1969; Khristianovich and Zheltov 1955), the Perkins-Kern-Nordgren (PKN) model (Nordgren 1972; Perkins and Kern 1961), and radial models (Abe et al. 1976), to more recent advanced numerical models, using discrete element methods (McClure et al. 2016; Sesetty and Ghassemi 2015; Zhao 2014), cohesive zone methods (Bryant et al. 2015; Ripudaman et al. 2016; Wang et al. 2016), and extended finite element methods (Dahi-Taleghani and Olson, 2011; Gordeliy and Peirce, 2013; Wang 2015; Wang 2016). Even though these advanced methods enable us to model fracture propagation with complex interactions under fully coupled physics, very few models have investigated the dynamic fracture closure process where the effect of fracture surface roughness and asperities cannot be neglected. The parallel plate idealization of fracture geometry is valid only for relatively smooth fracture surfaces and open fractures. This is the case during hydraulic fracture propagation when the fracturing fluid exerts sufficient pressure on the rock surface to prop it open and maintain a finite fracture aperture. However, as pressure declines inside the fracture after pump shut-in, the fracture will gradually close and the stress acting normal to the fracture plane results in fracture apertures approaching the scale of the surface roughness.

---

If the fracture faces are perfectly parallel and smooth, they will come into contact all at once when the fluid pressure inside the fracture declines to the far field stress, and the fracture is then mechanically and hydraulically closed. But there is abundant evidence to suggest that fractures retain their conductivity after the walls have come into contact on asperities (mechanical closure). Fractures retain a finite aperture after mechanical closure due to a mismatch of asperities in the fracture walls. van Dam et al. (2000) presented scaled laboratory experiments on hydraulic fracture closure behavior. Their work shows that the roughness of the fracture surfaces appears as a characteristic pattern of radial grooves (shown in **Fig.1**), and it is influenced by the externally applied stress. They also observed up to a15% residual aperture (compared to the maximum aperture during fracture propagation) long after shut-in. Fredd et al. (2000) demonstrated fracture surface asperities can provide residual fracture width and sufficient conductivity in the absence of proppants. Using sandstone cores from the East Texas Cotton Valley formation, sheared fracture surface asperities that had an average height of about 0.09 inches were observed. Warpinski et al. (1993) reported hydraulic fracture surface asperities of about 0.04 and 0.16 inches for nearly homogeneous sandstones and sandstones with coal and clay rich bedding planes, respectively. Zhang et al. (2014) conducted experiments on Barnett Shale samples and found that the surface topography of the displaced fracture can be altered because of rock failure, and the fracture surface exhibited parallel strips of crushed asperities. Field measurements (Warpinski et al. 2002) using a down-hole tiltmeter array indicated that the fracture closure process is a smooth, continuous one which often leaves 20%-30% residual fracture width, regardless of whether the injection fluid is water, linear-gel or cross-linked-gel.

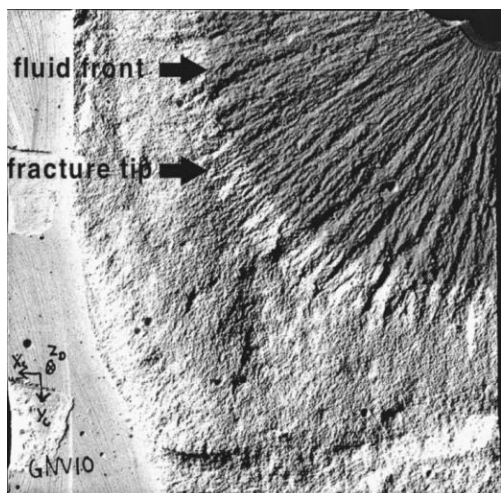

**Fig.1 Example of fracture surface roughness pattern (van Dam et al. 2000)**

Three-dimensional optical surface profiles of fracture surfaces obtained by Wu and Sharma (2017) shown that shale fracture surfaces can display asperities that range from 0.05 to 0.25 cm. If the fracture surface is exposed to acid, the surface roughness increases substantially as the calcite is dissolved away. The asperities on fracture surfaces act as pillars to keep the fracture open. Increasing the effective stress (as occurs during production) reduces the fracture aperture and its conductivity due to both elastic and plastic deformation of the asperities. Quantification of fracture closure is crucial to predicting the fracture conductivity of un-propped fractures and this cannot be done without a detailed study of the closure mechanisms.

In unconventional reservoirs, the connectivity, conductivity, and density of fracture networks formed by hydraulic fractures and induced natural fractures play a decisive role in determining production and its decline trend (Wang and Marongiu-Porcu 2015; Wang 2017a). In natural fractures, the most important properties that affect the fluid flow are the fracture width, normal stress, contact area and the roughness of the fracture surfaces. These properties are all interdependent and directly affect each other. In propped fractures, the stress distribution after closure can substantially influence the crushing or embedment of proppants and the resulting conductivity (Warpinski 2010). Those stress dependent fracture conductivity not only impact hydrocarbon production, but also disturb our rate transient analysis (RTA) of production data (Wang 2017b). In addition, the changes of fracture stiffness/compliance during closure can have a significant impact on the determination of in-situ stress using diagnostic fracture injection tests (DFITs) or flow back tests. van Dam et al. (2000) demonstrated that fracture surface roughness can enhance the fracture recession process by increasing the pressure gradient, and cause the pressure decline curve to deviate from the assumptions of linear elastic fracture mechanics. McClure et al. (2016) simulated pressure decline during DFITs by incorporating a residual aperture and contact stress into their numerical model. Their results show that fracture asperities can substantially alter the pressure decline curve and this poses a challenge in estimating minimum horizontal stress correctly. Raaen et al. (2001) proposed a "system stiffness approach" to interpret pressure data from DFITs and flow back tests. Their study indicates that the fracture stiffness increases during the closure process, and that the closure stress should be picked at the occurrence of mechanical closure, when asperities on the fracture surface begin to contact each other. Wang and Sharma (2017) presented a global DFIT model under a coherent mathematical framework that combines dynamic fracture closure process and fracture pressure dependent leak-off into a single material balance equation. Their work indicates that the minimum in-situ stress is overestimated if it is picked as the time at which the fracture compliance changes (McClure et al. 2016) and underestimated using conventional "tangent line" method (Barree et al. 2009) on a G-function or square root of time plot.

An extensive literature exists on the relationship between fracture aperture, conductivity, fluid pressure, and effective stress. Zangerl et al. (2008) compiled laboratory and in-situ experiments of fracture closure behavior for different rock samples and it shows that the effective normal stress and fracture normal closure (the displacement from zero effective normal stress condition) is highly non-linear, as shown in **Fig.2**. This relationship depends on many factors, such as rock mechanical properties, surface roughness, the distribution of asperities, etc. To quantify fracture closure behavior, numerous efforts have been made, include analytical models (Greenwood and Williamson, 1966; Gangi, 1978; Brown and Scholz, 1985; Cook, 1992; Adams and Nosonovsky, 2000; Myer, 1999) and numerical approaches (Pyrak-Nolte and Morris, 2000; Lanaro, 2000; Kamali and Pournik 2016) as well as experimental studies (Barton et al. 1985; Brown et al. 1998; Marache et al. 2008; Matsuki et al. 2008). However, all these studies focus on detailed investigations of the closure of fractures modeled as two parallel plates with rough surfaces. The impact of the fracture geometry and the evolution of the stress distribution on the fracture faces during fracture closure have not been quantified.

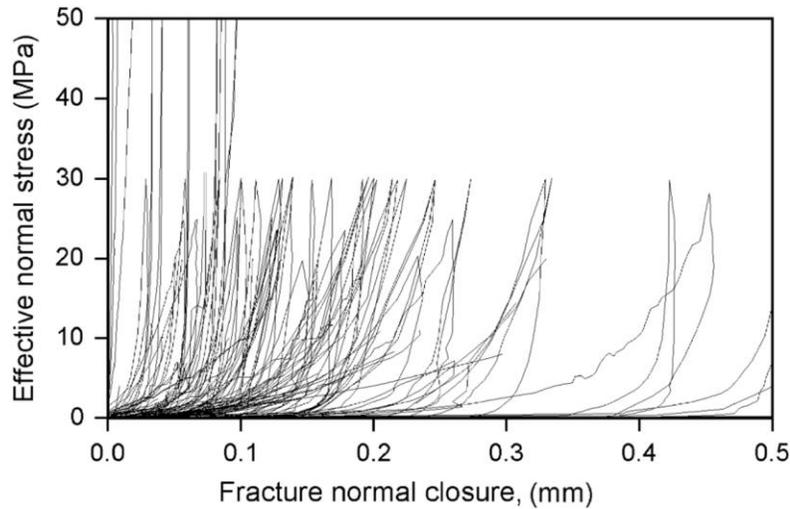

Fig.2 Compiled laboratory and in-situ experiments showing highly non-linear normal closure behavior (Modified from Zangerl et al. 2008)

In this article, we propose a method and general algorithms to model the dynamic behavior of hydraulic fracture closure on rough fracture surfaces and asperities. Analytic solutions that relate fracture width with net pressure/stress are coupled with a general contact law for three different fracture models (PKN, KGD and radial fracture geometry). The approaches presented in this study enable us to interpret fracture closure behavior as fluid pressure inside the fracture declines. The evolution of fracture aperture, total fracture volume and fracture stiffness/compliance can be determined given fracture geometry, rock properties and in-situ stress.

## 2. Fracture Width Distribution with Arbitrary Net Pressure/Stress

### 2. 1 PKN and KGD Fracture Geometry

England and Green (1963) showed that the width of a 2D fracture with any arbitrary pressure/stress profile is given by:

$$w_f(y) = \frac{16}{E'} \int_{|y|}^{a} \frac{F(\gamma) + \gamma G(\gamma)}{\sqrt{\gamma^2 - y^2}} d\gamma \qquad (1)$$

where $\gamma$ is a dummy variable, y is the distance to the middle of the fracture, a is the half fracture height (PKN model) or half fracture length (KGD model), as shown in **Fig.3**. The plane strain assumption is applied in the z-direction. $F(\gamma)$ and $G(\gamma)$ are functions to be determined from the pressure/stress distribution in the fracture. $E'$ is the plane strain Young's modulus and it is related to the Young's Modulus, E, and Poisson's Ratio, $\upsilon$ :

$$E' = \frac{E}{1-\upsilon^2} \qquad (2)$$

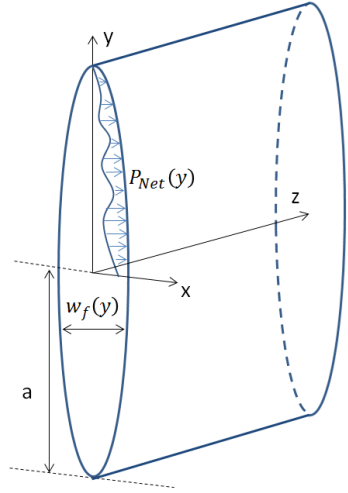

**Fig.3 Illustration of 2D planar fracture geometry.**

$P_{Net}(y)$ is the net pressure/stress distribution along the y-axis, which is the total pressure/stress acting on the fracture surfaces minus the far field stress in the fracture opening direction. To obtain $F(\gamma)$ and $G(\gamma)$, $P_{Net}(y)$ has to be divided into even and odd functions, such that $P_{Net}(y) = -f(y) - g(y)$, with $f(y)$ being even and $g(y)$ being odd. Then $F(\gamma)$ and $G(\gamma)$ are determined by:

$$F(\gamma) = -\frac{\gamma}{2\pi}\int_0^\gamma \frac{f(\beta)}{\sqrt{\gamma^2-\beta^2}}d\beta \qquad (3)$$

$$G(\gamma) = -\frac{1}{2\pi\gamma}\int_0^\gamma \frac{\beta g(\beta)}{\sqrt{\gamma^2-\beta^2}}d\beta \qquad (4)$$

where $\beta$ is a dummy variable. If the net pressure/stress distribution $P_{Net}(y)$ is symmetric to x-z plane, then $g(y)$ and $G(\gamma) = 0$. For example, assuming a uniform pressure $P_{Net}$ inside the fracture, $f(y) = -P_{Net}$, $g(y) = 0$, From Eq.(3):

$$F(\gamma) = -\frac{\gamma}{2\pi}\int_0^\gamma \frac{-P_{Net}}{\sqrt{\gamma^2-\beta^2}}d\beta = \frac{\gamma P_{Net}}{4} \qquad (5)$$

Substituting Eq.(5) into Equation (1), we have,

$$w_f(y) = \frac{16}{E'}\int_{|y|}^a \frac{\gamma P_{Net}}{4\sqrt{\gamma^2-y^2}}d\gamma = \frac{4P_{Net}}{E'}\sqrt{a^2-y^2} \qquad (6)$$

Equation (6) is the well-known equation for a static pressurized crack. If a is the fracture half-height, then Eq.(6) describes the fracture width distribution for the PKN fracture geometry, and if a is the fracture half-length, Eq.(6) describes the fracture width distribution for the KGD fracture geometry.

Depending on $P_{Net}(y)$, the existence of a solution to the above integral equations is not guaranteed. Numerical techniques have to be used when a closed form analytic solution is not available. To demonstrate the concept, the above example can be also solved numerically. Combine Eq.(1) and Eq.(3), with $f(y) = -P_{Net}$, $g(y) = 0$:

$$w_f(y) = \frac{8}{E'} \int_{|y|}^{a} \int_{0}^{\gamma} \frac{\gamma P_{Net}}{\pi\sqrt{\gamma^2 - \beta^2}\sqrt{\gamma^2 - y^2}} d\beta d\gamma \tag{7}$$

Notice that Eq.(7) in a double integral with dummy variables $\gamma$ and $\beta$. Consider only the upper half of the fracture ($y \geq 0$), the integration domain $\Omega$ is shown in the **Fig.4**. The real word problem in the x-y coordinates system is now projected onto an imaginary $\gamma$ - $\beta$ coordinate system.

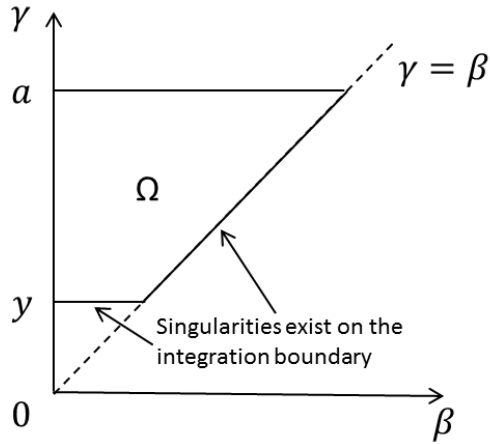

**Fig.4 Illustration of integration domain and singularities**

It can be observed that both the integrand $\frac{\gamma P_{Net}}{\pi\sqrt{\gamma^2-\beta^2}\sqrt{\gamma^2-y^2}}$ and the integration domain $\Omega$ changes with different values of $y$, also attention should be paid at the boundaries, where singularities reside on $\gamma = y$ and $\gamma = \beta$. To address this issue, special algorithms (Shampine 2008) are used to integrate Eq.(7) numerically. When an analytic form of $P_{Net}(y)$ is not available *a priori* (e.g., coupled with other physics) or $P_{Net}(y)$ is not a continuous function, numerical integration to calculate the fracture width has to be implemented in a discretized manner along the fracture height or length. **Fig.5** shows a fracture that is discretized into a number of *n* segments. In each segment, there exists a uniform net pressure/stress that acts on the fracture surfaces. Any net pressure/stress distribution can be approximated by representing it in this manner. The algorithms used to obtain the fracture width with a given arbitrary net pressure/stress distribution for PKN and KGD geometry are discussed in **Appendix-A**.

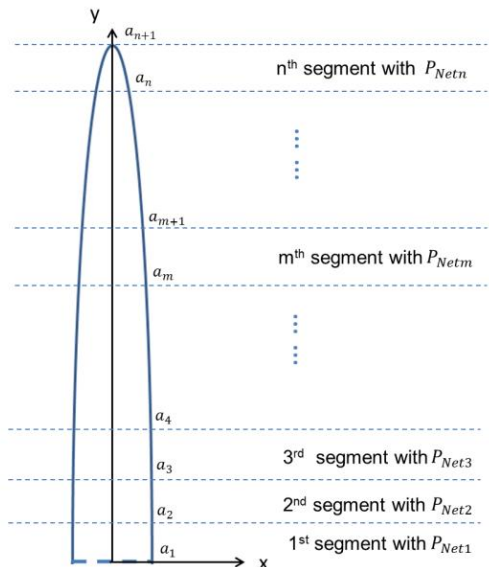

**Fig.5 Discretize fracture into n segments with different net pressure/stress**

To validate the algorithm presented in Appendix-A, a simple case is set up with a fracture half-height divided into three segments with equal length, and a different net pressure/stress is imposed on each segment. The resulting fracture width profiles are shown in **Fig.6**. As we can see, when the net pressure/stress is the same on all segments, an elliptical fracture width profile is obtained, which is also validated by Eq.(6). When the net pressure/stress varies along different segments, the fracture width profile will depart from the smooth, elliptical curve. And $P_{Netn}$ has more impact on the overall width distribution when it is imposed close to fracture center. From the perspective of a fluid driven hydraulic fracture, different net pressures/stresses can stem from the contact of rough fracture surfaces or non-uniform horizontal stress in different rock layers.

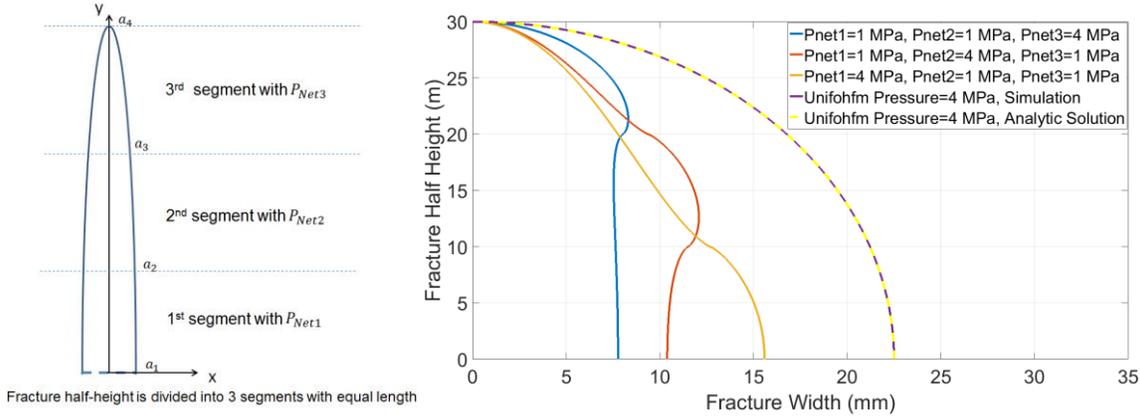

**Fig.6 Fracture width distribution along fracture half-height across three segments: *a*=30 m, E=20 GPa, ν=0.25.**

## 2. 2 Radial Fracture Geometry

The radial crack opening due to an excess pressure distribution in a linearly elastic material of infinite extent can be expressed as (Sneddon and Lowengrub 1969):

$$w_f(r_D) = \frac{8R}{\pi E'} \int_{r_D}^{1} \frac{du}{\sqrt{u^2 - r_D^2}} \int_0^u \frac{s P_{Net}(s) ds}{\sqrt{u^2 - s^2}} \tag{8}$$

where $R_f$ is the length of fracture radius, $u$ and $s$ are dummy variables and $r_D$ is the normalized radius, which is defined as

$$r_D = \frac{r}{R_f} \tag{9}$$

where $r$ is the local fracture radius. For a radius crack with a uniform pressure distribution $P_{Net}$, Eq.(8) can be integrated directly to get an analytical solution for the fracture width profile:

$$w_f(r_D) = \frac{8 R_f P_{Net}}{\pi E'} \sqrt{1 - r_D^2} \tag{10}$$

Barr (1991) proved that the double integral of Eq.(8) can be transformed into a single integral with the aid of elliptic integrals:

$$w_f(r_D) = \frac{8R_f}{\pi E'} \left\{ \int_0^{r_D} s P_{Net}(s) \left[ \frac{1}{r_D} F\left( \sin^{-1}\sqrt{\frac{1 - r_D^2}{1 - s^2}}, \frac{s}{r_D} \right) \right] ds + \int_{r_D}^1 P_{Net}(s) \left[ F\left( \sin^{-1}\sqrt{\frac{1 - s^2}{1 - r_D^2}}, \frac{r_D}{s} \right) \right] ds \right\} \tag{11}$$

$F(\varphi, k)$ is the elliptic integral of the first kind, which is defined as:

$$F(\varphi, k) = \int_0^{\varphi} \frac{d\theta}{\sqrt{1 - k^2 \sin^2 \theta}} \tag{12}$$

where $\varphi$ and $k$ are input parameters for elliptic integral, and $\theta$ is a dummy variable.

**Fig.7** shows a radial fracture discretized into n segments. In each segment, there exists a uniform net pressure/stress that acts on the opposite fracture surface. If $n$ is large enough, the net pressure/stress distribution can be approximated to any distribution of $P_{Net}(r)$. Details on how to obtain the fracture width with a given arbitrary net pressure/stress distribution for a radial geometry are presented in **Appendix-B**.

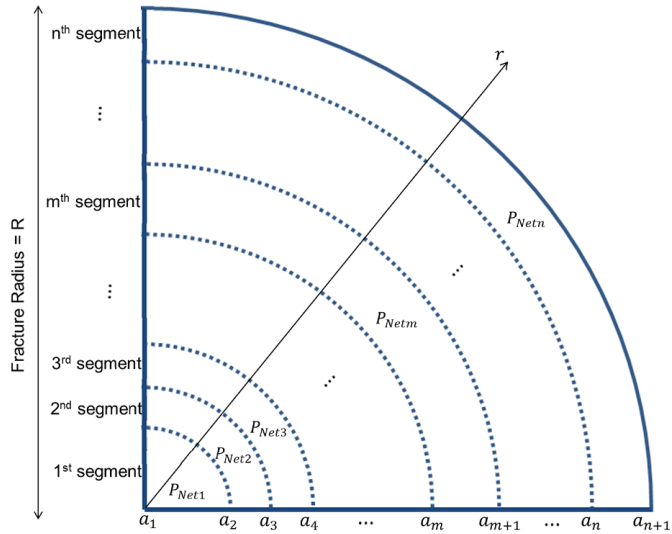

**Fig.7 Discretized radial fracture (n segments) with different net pressure/stress acting along the radial direction.**

To validate the algorithm presented in Appendix-B, a radial fracture geometry is set up with a fracture radius divided into three segments with equal length, and a different net pressure/stress is imposed on each segment. The resulting fracture width profiles are shown in Fig.8. As we can see, when the net pressure/stress is the same on all segments, it reproduces the same fracture width distribution as predicted by Eq.(10). And when the net pressure/stress varies along different segments, the fracture width within a certain segment can be substantially impacted by the load imposed on other segment.

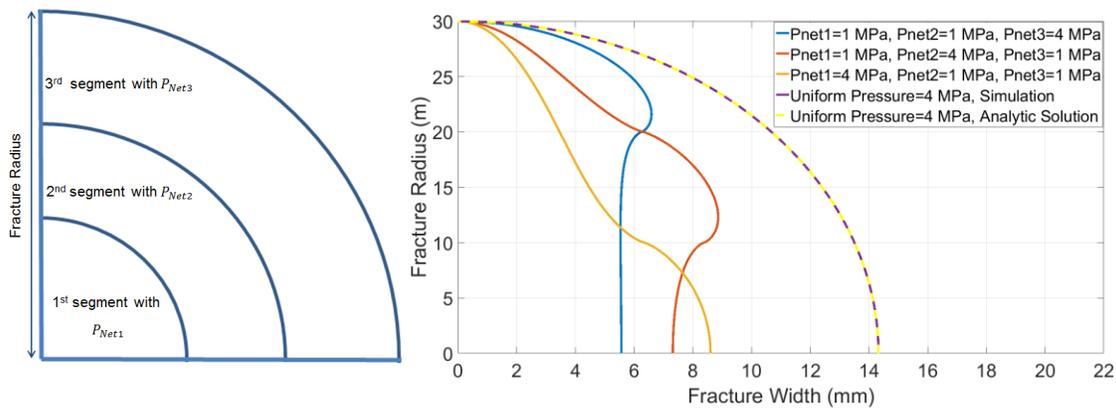

**Fig.8 Fracture width distribution along fracture radius across three segments: $R_f$=30 m, E=20 GPa, v=0.25.**

Up to this point, we have developed a method and proposed general algorithms to determine the fracture width profile for PKN, KGD and radial fracture geometries, given an arbitrary distribution of net pressure/stress. Since both PKN and KGD models assume plane strain and only differ in the direction of plane strain, one only needs to substitute $a$ with fracture half height for the PKN model and the fracture half-length for the KGD model

### 3. Couple Contact Stress with Fracture Width and Pressure

If the fracture surface is perfectly smooth, according to Eq.(6) and Eq.(10), the entire fracture surface area will get into contact all at once when the internal pressure drops to zero. However, as discussed in previously, depending on material properties, heterogeneity and loading conditions, the generated crack or fracture surface is always rough with asperities distributed across the surface area. So fractures can retain a finite aperture after mechanical closure due to a mismatch of asperities in the fracture walls, this is especially the case for subsurface fractures that are stimulated in unconventional and geothermal reservoirs. Detailed measurement and modeling of surface roughness and mechanical properties of asperities for every fracture encountered is not practical and, even impossible for large scale fractures, so it is desirable to upscale the influence of surface microscopic structure to macroscopic empirical relationships (i.e., a contact law) that relates fracture width and the associated contact stress. As an example we have chosen to use a relationship that Willis-Richards et al. (1996) used for fracture aperture, based on the work of Barton et al. (1985):

$$\sigma_c = \frac{\sigma_{ref}}{9}\left(\frac{w_0}{w_f} - 1\right) \text{ for } w_f \leq w_0 \tag{13}$$

where $w_f$ is the fracture aperture and, $w_0$ is the contact width that determined by the tallest asperities, it is the fracture aperture when the contact normal stress is equal to zero, $\sigma_c$ is the contact normal stress on the fracture, and $\sigma_{ref}$ is a contact reference stress, which denotes the effective normal stress at which the aperture is reduced by 90%. Essentially, this equation can be used to quantify the relationship between effective normal stress and fracture width by adjusting $w_0$ and $\sigma_{ref}$ to match measured values (e.g., Fig.2). The influence of all the rock properties, asperities patterns, density, distribution, etc., are up-scaled into these two contact parameters.

In a fluid driven fracture, the net pressure/stress is determined by:

$$P_{Net}(y) = P_f(y) + \sigma_c(y) - \sigma_{hmin}(y) \tag{14}$$

where $P_f$ is the fluid pressure inside the fracture, $\sigma_{hmin}$ is the far field stress perpendicular to the fracture surface, i.e. the minimum horizontal stress. Note that the coordinate variable $y$ here means that pressure, far field stress and contact stress have spatial distributions, and can vary along the direction of fracture height (PKN geometry), length (KGD geometry) or radius (radial geometry).

In the previous section, we have proposed and validated our general algorithm to calculate the fracture width, $w_f(y)$, given an arbitrary net pressure/stress distribution of $P_{Net}(y)$. Now, if the fracture closes on rough surfaces and asperities, then the net pressure/stress, $P_{Net}(y)$, is impacted by the contact stress, $\sigma_c(y)$, which itself is a function of fracture width distribution. Therefore, the net pressure/stress distribution of $P_{Net}(y)$ and fracture width, $w_f(y)$, have to be solved simultaneously. Because both the net pressure/stress distribution and fracture width distribution are discretized, they can be only solved iteratively. **Fig.9** shows the numerical scheme to solve the net pressure/pressure and fracture width distribution.

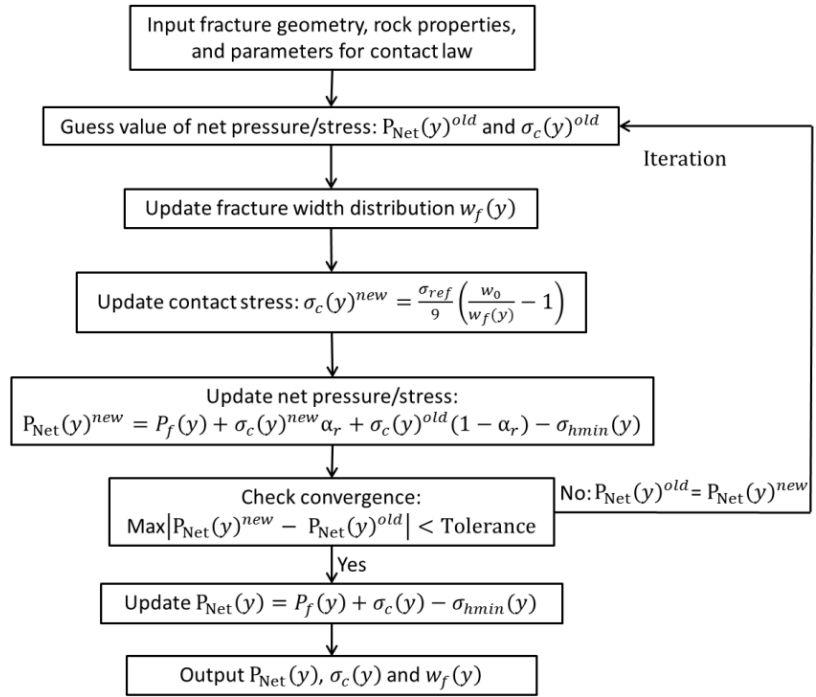

Fig.9 Solution flow diagram for solving coupled net pressure/stress and fracture width distribution

During the first step, if the contact stress $\sigma_c(y)$ is not known, $\sigma_c(y)$ is assumed to be zero, and the resulting fracture width is smaller than the actual width if some portion of the fracture tip is already in contact. Based on this first attempt of fracture width estimation, we can calculate the corresponding contact stress from Eq.(13). Then the new net pressure/stress distribution is updated within the iteration:

$$P_{Net}(y)^{new} = P_f(y) + \sigma_c(y)^{new}\alpha_r + \sigma_c(y)^{old}(1-\alpha_r) - \sigma_{hmin}(y) \tag{15}$$

In this Picard iteration scheme, $\alpha_r$ is a relaxation factor that is less than or equal to 1.0. It controls how much the net pressure/stress should be corrected within iterations. Smaller $\alpha_r$ can avoid over-correction and improve convergence stability, but more iterations may be required with small value of $\alpha_r$. When the changes in net pressure/stress between iterations is less than the tolerance value, the estimation of contact stress, $\sigma_c(y)$, is converged ($\sigma_c(y)^{new} \approx \sigma_c(y)^{old}$) and the final net pressure/stress distribution can be recovered from Eq.(14). This proposed iteration method is general and can be applied to any fracture model. However, to model fracture closure process more efficient, some additional techniques are applied. The fluid pressure has to decline from a value that is higher than the far field stress avoid negative net pressure/stress. To reduce the number of iterations, the net pressure/stress for the current iteration step is obtained from the converged value of the last iteration. In general, surface contact problems are highly nonlinear, as reflected by Eq.(13) and Fig.2, where the resulting contact stress increases exponentially as fracture width/aperture decreases. This poses a challenge for the numerical scheme, because when a small change in fracture width leads to a dramatic increase in contact stress, the numerical iteration itself becomes unstable, unless a small relaxation factor and fluid pressure interval is used. Initially, when the fluid pressure inside fracture is relative high, small changes in fracture width only lead to moderate increase in contact stress, so a large relaxation factor can be used to reduce the iteration number and computation time. In order to achieve convergence in a reassemble time frame, a dynamic relaxation factor is used. During the simulation, when fluid pressure drops to a point where the maximum iteration number is reached, the relaxation factor is automatically cut to half of its original value. And when the relaxation factor is decreased the tolerance should also decrease accordingly.

## 4. Dynamic Fracture Closure Process

### 4.1 PKN Fracture Geometry

Assuming a base case for PKN fracture geometry with 10 m fracture height, 50 m fracture half-length, and 35 MPa minimum far-field in-situ stress that perpendicular to fracture surface, and the Young's modulus is 20 GPa, Poisson's ratio is 0.25, $w_0$ is 2 mm, and $\sigma_{ref}$ is 5 MPa. The evolution of the fracture width profile and the corresponding contact stress distribution can be determined, as the fluid pressure inside the fracture gradually declines. **Fig.10** shows the fracture width distribution at different fracturing fluid pressure intervals, which is assumes uniform inside fracture. To demonstrate the impact of fracture roughness and surface asperities on fracture closure behavior, the case without surface asperities (fracture surface is completely smooth) is also included. Fracture closure on rough walls is calculated based the algorithm presented in the previous section and fracture closure on a smooth surface is determined using Eq.(6). The result shows that at relative high fracturing fluid pressure, the fracture asperities have negligible impact on fracture width distribution, and as pressure declines, the role of asperity contact start to manifest itself by making fracture resilient to further closure. When fluid pressure drops below minimum in-situ stress of 35 MPa, fracture width collapses to zero for the case of smooth fracture surfaces, but still retain a residual fracture width when surface roughness exists. **Fig.11** shows the corresponding contact stress distribution at different fracturing fluid pressure interval when fracture closes on asperities. The result indicates that the contact stress always concentrates at the tip of the fracture, where the contact stress is much higher than that in the middle of the fracture. We can also see that the fracture surfaces do not contact each other like parallel plates. In fact, the fracture closes on rough surfaces starting from the tip, and closes progressively all the way from the edges to the center of fracture. This explains why the fracture asperities have less influence on fracture closure behavior when fracturing fluid is relative high. Because at high fluid pressure, the overall fracture width is still larger than the contact width, $w_{f0}$, during this and the contact stress only impacts the very tip of the fracture. However, as fluid pressure continues to decline, more and more of the fracture surfaces come into contact, and this ultimately changes the subsequent fracture closure behavior. At lower fluid pressures, contact stresses start to counter-balance the in-situ stress and the fracture itself becomes stiffer, and less compliant to further pressure drops. The moment when all fracture surfaces come into contact on asperities and the contact stress becomes non-zero on the entire fracture surface, the fracture is completely mechanically closed and this mechanical close stress is higher than the fracture minimum in-situ stress.

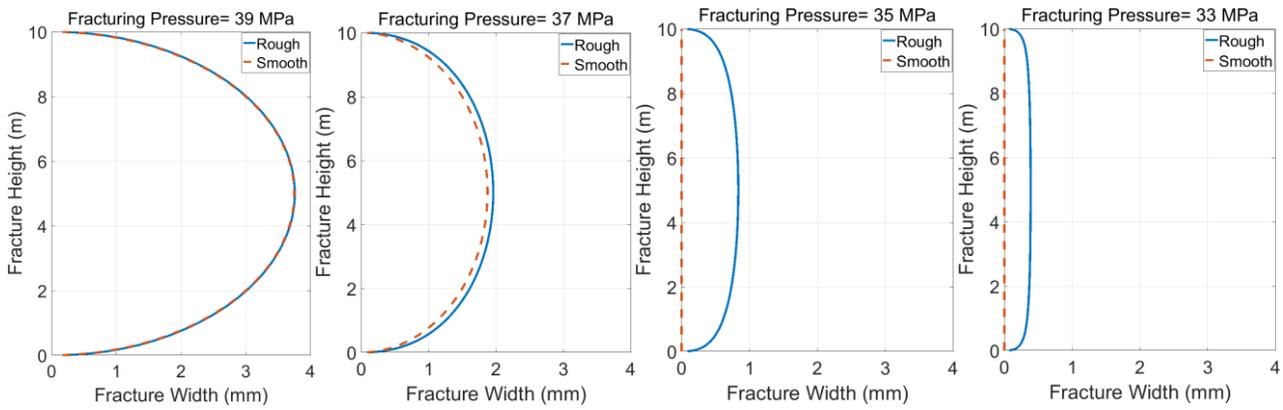

**Fig.10 Fracture width evolution with and without asperities at different fluid pressure for a PKN geometry**

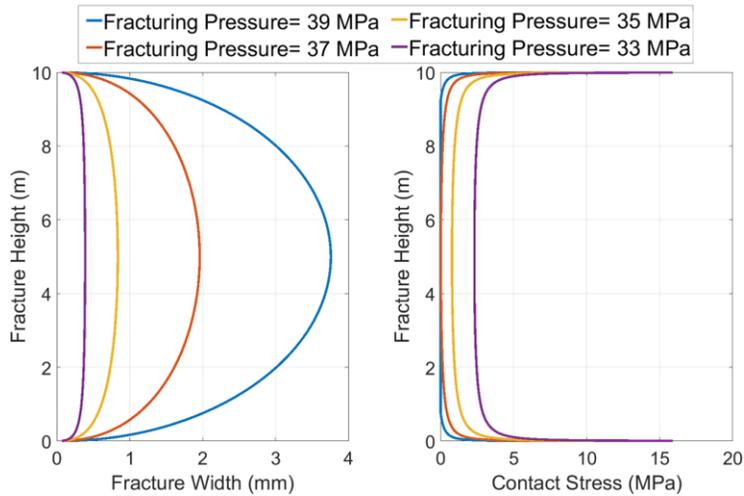

**Fig.11 Fracture width and corresponding contact stress distribution at different fluid pressure for a PKN geometry**

**Fig.12** shows fracture volume evolution as a function of fluid pressure. As can be seen, when the fluid pressure inside the fracture is relatively high, the fracture volume declines linearly with pressure. However, as the pressure declines to a certain level the fracture volume and pressure departs from a linear relationship.

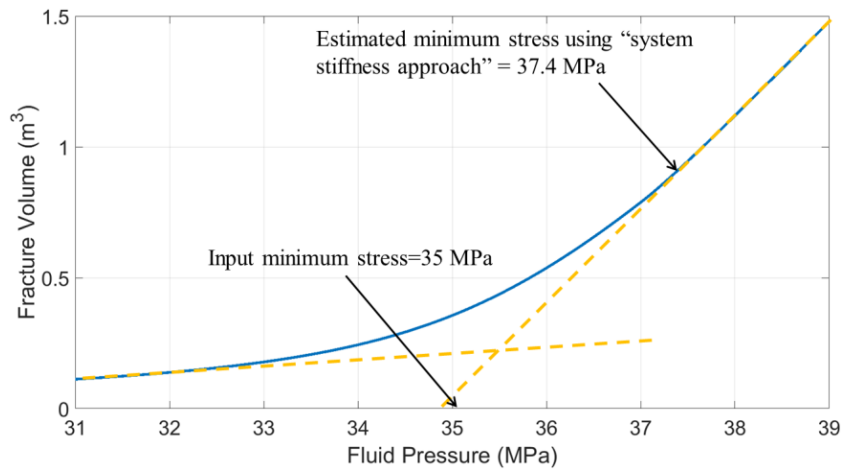

**Fig.12 Fracture volume evolution as fluid pressure declines for a PKN geometry**

Here we introduce a very important concept of fracture stiffness, which can be defined as:

$$S_f = \frac{A_f \, dP_f}{dV_f} \qquad (16)$$

where $A_f$ is the fracture surface area, and $V_f$ is fracture volume. For a fracture with constant surface area $A_f$, Eq.(16) can be re-written as

$$S_f = \frac{A_f dP_f}{d(A_f \overline{w_f})} = \frac{dP_f}{d\overline{w_f}} \tag{17}$$

where $\overline{w_f}$ is the average fracture. Essentially, fracture stiffness represents fracture compressibility. The higher the value of the fracture stiffness, the smaller the changes in fracture volume or average fracture width for unit pressure drop inside the fracture. The compressibility of fracture can also be represented by fracture compliance, which is the reciprocal of fracture stiffness. For a fracture with perfectly smooth walls, the fracture stiffness during closure is a constant value and can be determined analytically, as shown in **Table 1**.

| Fracture Geometry | PKN | KGD | Radial |
|---|---|---|---|
| $S_f$ | $\dfrac{2E'}{\pi h_f}$ | $\dfrac{E'}{\pi x_f}$ | $\dfrac{3\pi E'}{16 R_f}$ |

**Table 1-Fracture stiffness expressions for different fracture geometry models**

The "system stiffness approach" proposed by Raaen et al. (2001), Gederaas and Raaen (2009) suggested that the minimum stress should be picked at the moment where stiffness starts to changes when interpreting pressure data from DFITs and pump in/flow back tests. However, as reflected in Fig.12, the estimated minimum stress from the "system stiffness approach" is 37.4 MPa, which is 2.4 MPa higher than our input minimum stress. This is because the fracture closure process was modeled as the closure of two parallel, smooth plates, and under this premise, the fracture stiffness only changes at the moment when all fracture faces come into contact simultaneously and fracturing fluid pressure drops to minimum stress. However, fracture closure is a dynamic process where the mechanical closure initially occurs at the fracture edges and moves progressively toward fracture center/perforation, so the fracture stiffness changes over time as fluid pressure declines, even before fracturing fluid pressure drops to minimum stress.

The relationship between $P_f$ and $V_f$ in Fig.12 can be used to calculate the pressure dependent fracture stiffness during closure and the result is shown in **Fig.13**. Fracture stiffness only remains constant at relative high fracture pressures and starts to increase noticeably when fluid pressure drops to 37.4 MPa. So constant fracture stiffness calculation based on Table 1 is only applicable when fracturing fluid is relative high and the influence of asperity contact at fracture edges is negligible. McClure et al. (2016) proposed a "compliance method" for picking minimum stress on the G-function plot, where the minimum stress is picked at the point where the fracture stiffness start to increase. Similar to the "system stiffness approach", their method can overestimate the minimum in-situ stress because fracture stiffness increase is a dynamic process during closure as more and more fracture surface area gets into contact.

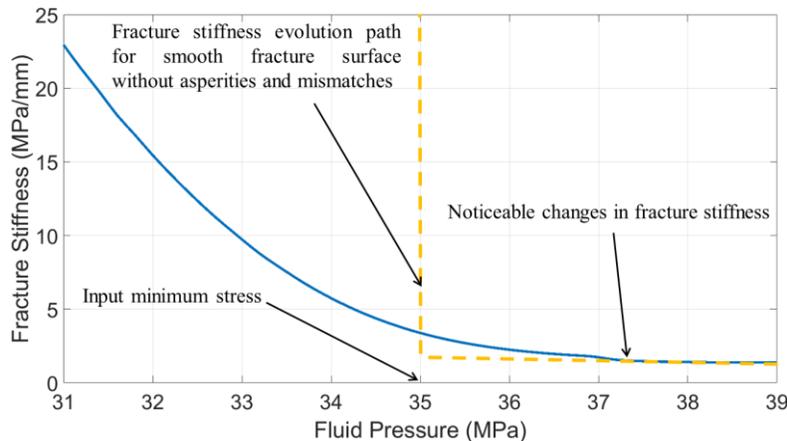

**Fig.13 Fracture stiffness evolution as fluid pressure declines for a PKN geometry.**

There is compelling field evidence to indicate that hydraulic fracture closure occurs on rough fracture surfaces in the field. **Fig.14** shows the normalized tiltmeter response plotted against wellbore pressure during the shut-in period of 2B well at three different stations from the GRI/DOE M-site. Each tiltmeter response is normalized by dividing by the maximum tilt (displacement, also an indication of fracture width) measured at that instrument during the test. The tiltmeter demonstrates that soon after shut-in, the measured displacement declines linearly with pressure (roughly constant fracture stiffness). If the fracture surface area remains constant during shut-in, then the fracture volume is proportional to the average displacement and should also decline linearly with the pressure within this period. After the wellbore pressure declines to a certain level, the measured displacement vs pressure departs from a linear relationship. This field measurement is consistent with the general trend shown in Fig.12. As pressure continues declining, more and more fracture surface area comes into contact with the rough fracture surfaces and asperities, and the fracture stiffness increases gradually. Even though these data were measured from the end of injection to weeks of shut-in, we can still observe the existence of a residual fracture width that is supported by surface asperities and mismatches even after the fracturing fluid pressure drops below the minimum in-situ stress.

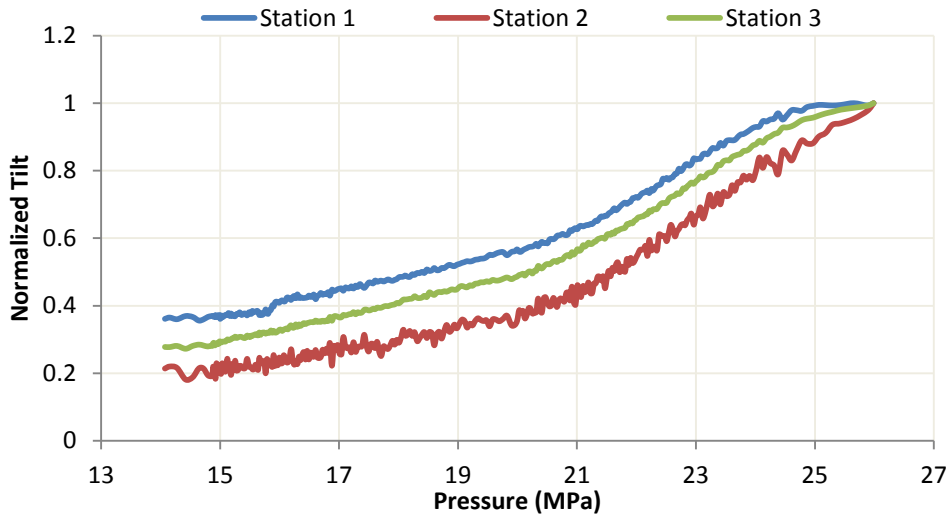

Fig.14 Normalized tiltmeter data from shut-in of 2B from the GRI/DOE M-site. Data Courtesy of Norm Warpinski

Next, we examine how contact width, $w_0$, and contact reference stress, $\sigma_{ref}$, impact dynamic fracture closure behavior. **Fig.15** shows fracture width and the corresponding contact stress for different values of $w_0$ with $\sigma_{ref}$ of 5 MPa using Eq.(13). As expected, when the fracture width is larger than or equals to the contact width, the contact stress is zero. However, when the fracture width is smaller than the contact width, the contact stress and fracture width follows a hyperbolic relationship. This trend resembles Fig.2, where the contact stress and fracture width relationship is highly non-linear.

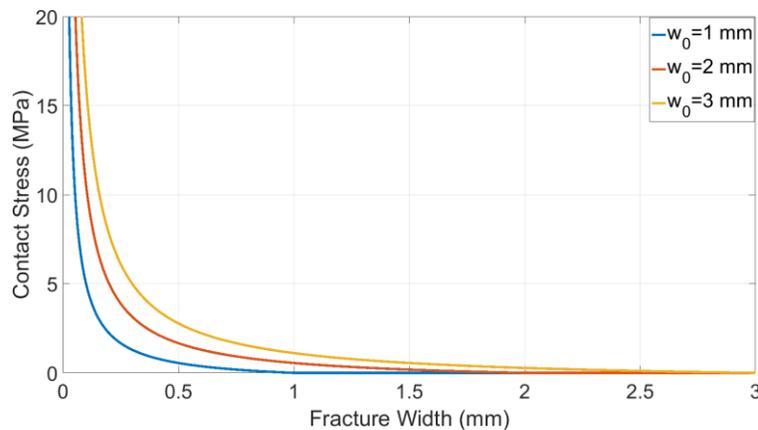

Fig.15 The relationship between contact stress and fracture width for different $w_0$

**Fig.16** and **Fig.17** shows the evolution of fracture volume and fracture stiffness for different contact width $w_0$, while other input parameters remain the same as the base case of PKN geometry. As can be observed, when the pressure is relative high (above 38.5 MPa), the fracture volume and fluid pressure follows a linear relationship and the fracture stiffness remains roughly constant for all cases. However, as pressure continues to decline, lager contact width leads to sooner departure of the linear relationship and fracture stiffness increases at a higher pressure. This is because the rough fracture walls will come into contact sooner if the contact width is larger, so the noticeable changes of fracture stiffness occur earlier when the contact width is larger. We can also observe that when the contact width is larger, the increase in fracture stiffness is more gradual and smooth.

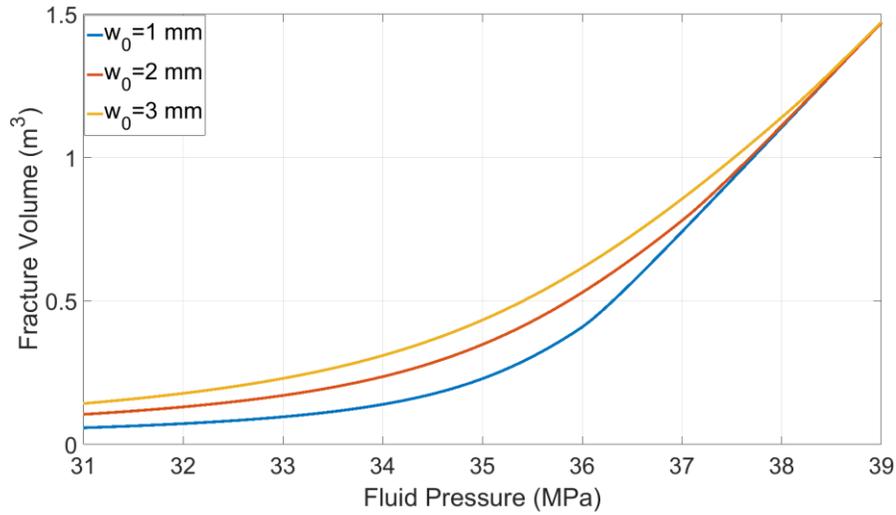

**Fig.16** Fracture volume evolution for different $w_0$ with a PKN geometry

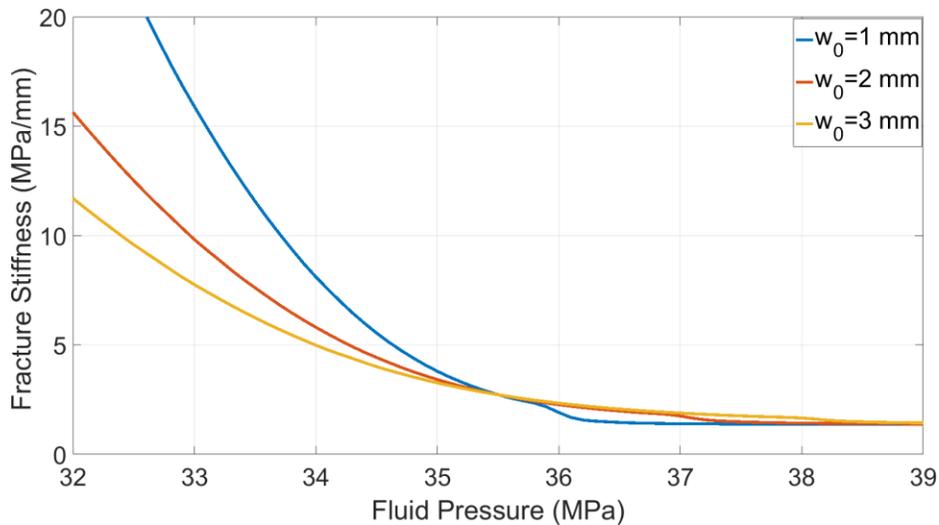

**Fig.17** Fracture stiffness evolution for different $w_0$ with a PKN geometry

**Fig.18** shows fracture width and corresponding contact stress for different values of $\sigma_{ref}$ with $w_0$ of 2mm using Eq.(13). It can be observed that for the same contact width, the higher the contact reference stress, the more rapid the increase of contact stress as the fracture width shrinks. Physically, the contact reference stress represents how hard and strong the fracture surface asperities are.

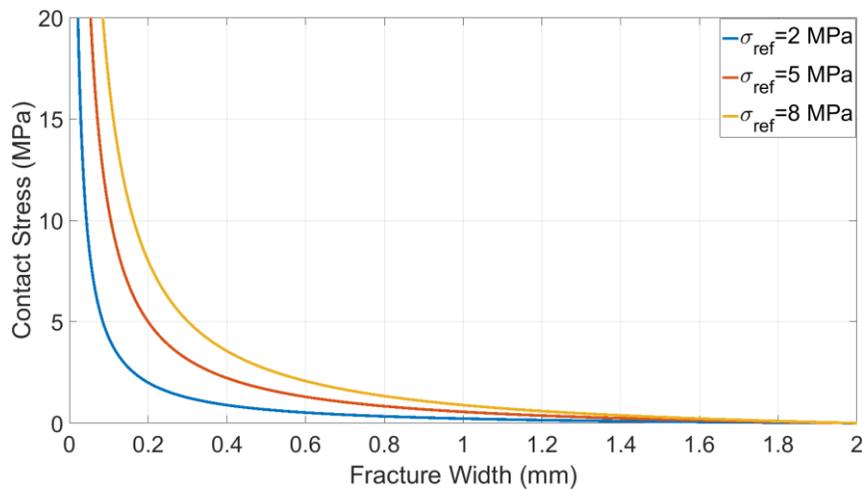

**Fig.18** The relationship between contact stress and fracture width for different $\sigma_{ref}$

**Fig.19** and **Fig.20** shows the evolution of fracture volume and fracture stiffness for different contact reference stress $\sigma_{ref}$, while other input parameters remain the same as the base case of PKN geometry. The results reveal that contact reference stress does not have much impact on the pressure at which the fracture volume and pressure departs from a linear relationship and fracture stiffness starts to changes noticeably, but it does impact the fracture stiffness evolution. The lower the contact reference stress, the more smooth the change in fracture stiffness as pressure declines, and lower the contact reference stress also leads to smaller residual fracture volume.

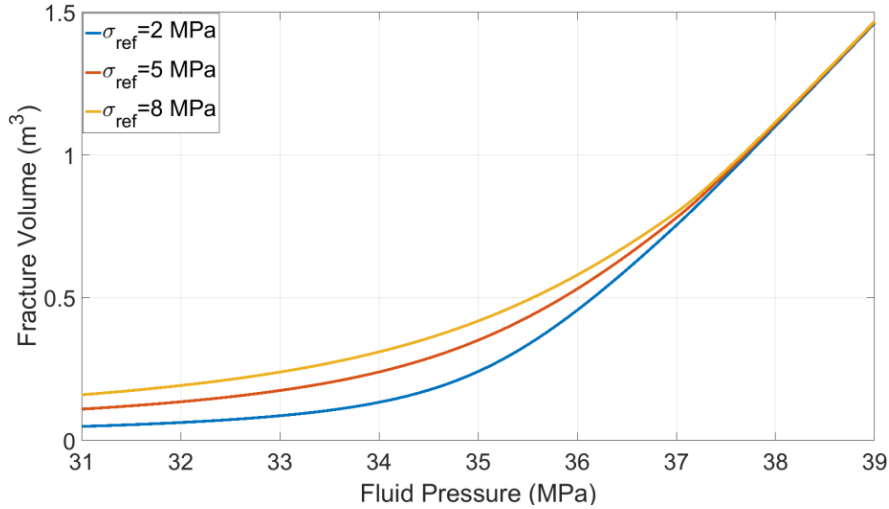

**Fig.19 Fracture volume evolution for different $\sigma_{ref}$ with a PKN geometry**

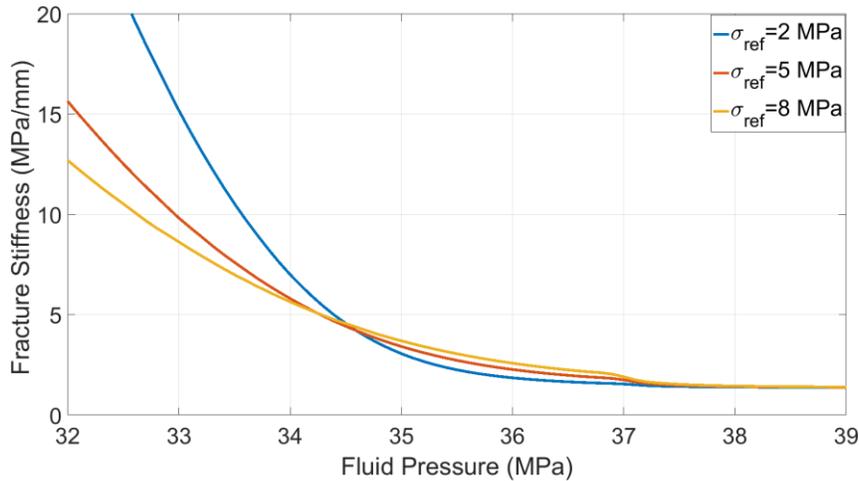

**Fig.20 Fracture stiffness evolution for different $\sigma_{ref}$ with a PKN geometry**

Even though this section primarily focused on the fracture closure behavior of a PKN type fracture, where the plane strain condition is assumed in the direction of fracture length, the same procedure can be applied to a KGD type fracture, where the plane strain condition is assumed in the direction of fracture height. Because similar results are obtained for KGD type fracture by replacing fracture half-height with fracture half-length when calculated fracture width distribution (Appendix-A), a detailed discussion of KGD type fracture closure behavior is not presented here. But it should be mentioned that, for a PKN fracture, fracture closes starting from the fracture top and bottom edges and for a KGD fracture, the fracture closes starting from the fracture tip.

### 4.2 Radial Fracture Geometry

The algorism for modeling non-local fracture closure behavior that presented in Fig.9 is a general approach. To model radial fracture closure behavior, the fracture width distribution under arbitrary normal load will be calculated using Appendix-B. Assuming a radial fracture with a radius of 10 m, and all the other input parameters remain the same as the previous PKN base case, **Fig.21** and **Fig.22** shows the evolution of fracture volume and fracture stiffness for different contact width $w_0$. Similar to the results of PKN fracture geometry, lager contact width leads to sooner departure of the linear relationship between fracture volume and fluid pressure, and more gradual and smooth increase in fracture stiffness.

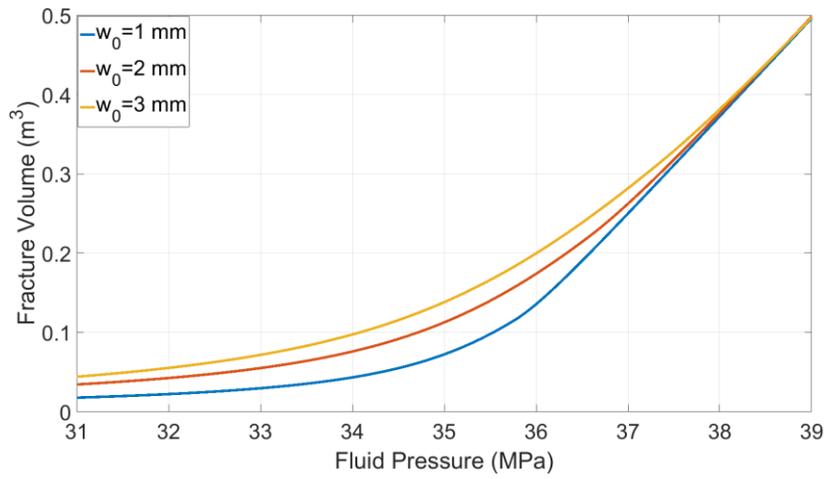

**Fig.21** Fracture volume evolution for different $w_0$ with a radial geometry

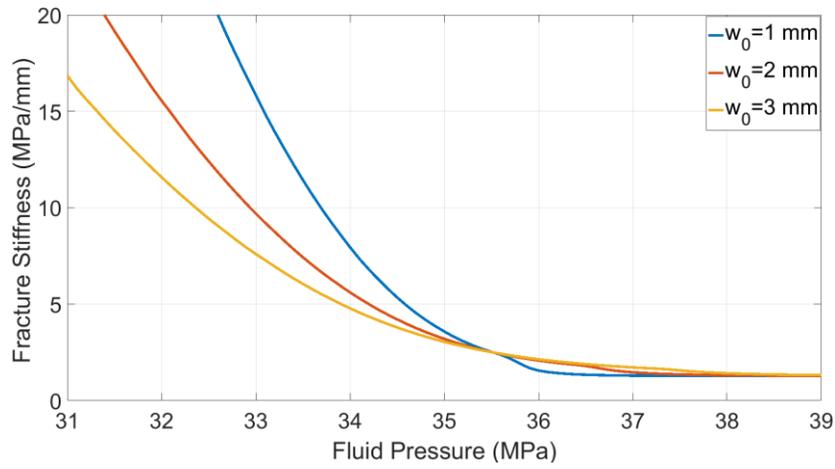

**Fig.22** Fracture stiffness evolution for different $w_0$ with a radial geometry

**Fig.23** and **Fig.24** shows the evolution of fracture volume and fracture stiffness for different contact reference stress $\sigma_{ref}$. Again, we can observe that contact reference stress strongly affect the stiffness evolution after a certain portion of fracture surface already get into contact, but it has less influence on at what pressure the contacted asperities begin to influence the overall fracture closure behavior noticeably.

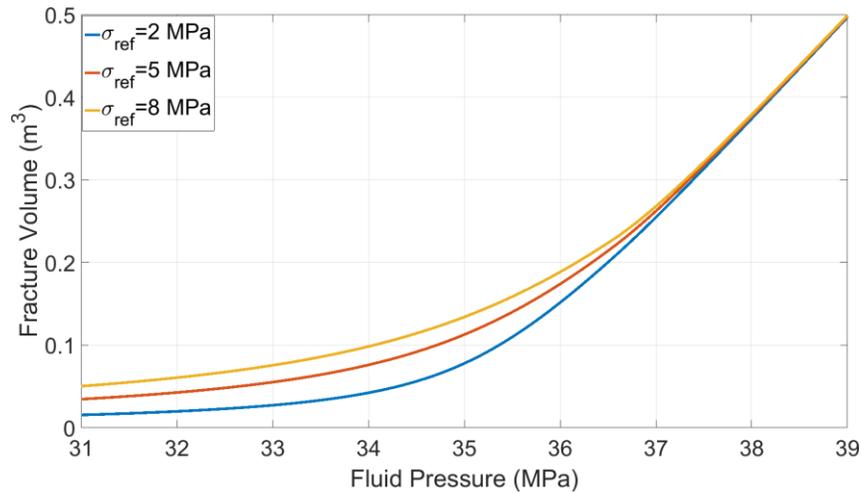

**Fig.23** Fracture volume evolution for different $\sigma_{ref}$ with a radial geometry

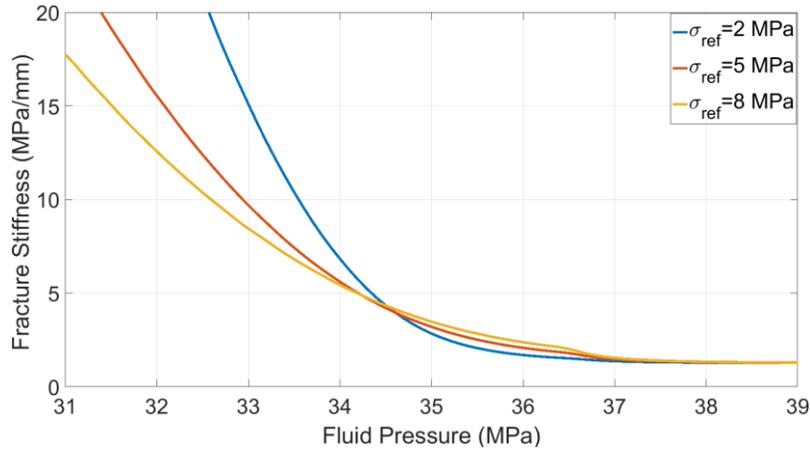

**Fig.24 Fracture stiffness evolution for different $\sigma_{ref}$ with a radial geometry**

From all the above simulated cases, it is clear that fracture closure behavior is substantially impacted by asperities and mismatches of rough fracture walls. Because fracture closes progressively from fracture edges towards the center of fracture or perforation, the fracture stiffness increases gradually as more and more fracture surface area come into contact, even before fracturing fluid pressure drops to the minimum in-situ stress. Therefore, any approaches trying to measure the minimum in-situ stress based on occurrence of changes in fracture stiffness will overestimate the minimum in-situ stress.

### 5. Conclusions and Discussion

The quantification of fracture closure is crucial in many aspects of petroleum engineering, such as predicting the performance of acid fracturing, evaluating the influence of pressure dependent conductivity of fracture networks in unconventional reservoirs, the placement of proppants after shut-in and estimating the minimum horizontal stress using diagnostic fracture injection tests (DFITs) or pump in/flow back tests. To model the dynamic fracture closure process it is essential to consider both the presence of surface asperities on the fracture surface and the shape of the original fracture. Past models that have investigated fracture closure mechanisms, have either neglected the role of fracture surface asperities or assumed that the fracture faces are two parallel plates and neglected the initial fracture geometry.

In this article, we propose a general method to model the dynamic behavior of hydraulic fracture closure for fractures that have realistic fracture width profiles and rough surfaces and asperities. Analytic solutions from linear elastic fracture mechanics for three fracture models (PKN, KGD and radial fracture geometry) are coupled with a general contact law. The approaches presented in this study enable us to model non-local fracture closure behavior and have a better understanding of the evolution of fracture aperture, fracture volume and fracture stiffness as the fluid pressure inside a fracture declines. This study also reveals that the fracture surface roughness and asperities do impact the dynamic fracture closure behavior and the determination of closure stress. Mechanical closure (contact of rough fracture faces) occurs at the fracture edges initially and then moves progressively to the perforation or the center of fracture, and so the closure stress should not be picked at the first occurrence of mechanical closure as previous studies suggest. Because the dynamic fracture closure model and the algorithm presented in this article is general it can be easily extended to couple with other mechanisms of interests, such as fracture propagation and fluid leak-off, etc. Based on the results presented in this paper, further research efforts are needed to develop a robust method for estimating in-situ stress and fracture stiffness/compliance from fracture injection tests.

### Nomenclature

$a$     = Half fracture height (PKN model) or half fracture length (KGD model), $m$.
$a_m$    = Half fracture height (PKN model) or half fracture length (KGD model) or local fracture radius (Radial model) at the start point of the $m^{th}$ segment along a discretized fracture, $m$.
$A_f$    = Fracture surface area of one face of one wing, $m.^2$
$E$     = Young's modulus, Pa
$E'$    = Plane strain modulus, Pa
$h_f$    = Fracture height, $m$.
$I$     = Integration operator for PKN and KGD model
$P_f$    = Fluid pressure inside fracture, Pa
$P_{Net}$   = Net pressure/stress, Pa
$P_{Netm}$   = Net pressure/stress at the $m^{th}$ segment along a discretized fracture, Pa

| | | |
|---|---|---|
| $r$ | = | Local fracture radius, $m$. |
| $r_D$ | = | Normalized radius |
| $R_f$ | = | Fracture radius, $m$. |
| $R1$ | = | First integration operator for Radial model |
| $R2$ | = | Second integration operator for Radial model |
| $s$ | = | Dummy variable |
| $S_f$ | = | Fracture stiffness, Pa/$m$. |
| $u$ | = | Dummy variable |
| $V_f$ | = | Fracture volume of one wing, $m.^3$ |
| $w_0$ | = | Contact width, $m$. |
| $w_f$ | = | Local fracture width, $m$. |
| $\overline{w_f}$ | = | Average fracture width, $m$. |
| $w_0$ | = | Contact width, $m$. |
| $x_f$ | = | Fracture half-length, $m$. |
| $\alpha_r$ | = | Relaxation factor |
| $\beta$ | = | Dummy variable |
| $\gamma$ | = | Dummy variable |
| $\theta$ | = | Dummy variable |
| $\upsilon$ | = | Poisson's ratio |
| $\sigma_c$ | = | Contact stress, Pa |
| $\sigma_{hmin}$ | = | minimum horizontal stress, Pa |
| $\sigma_{ref}$ | = | Effective normal stress at which aperture is reduced by 90%, Pa |

## Appendix-A: Determine Fracture Width with Arbitrary Normal Load for PKN and KGD Fracture Geometry

In the following derivation, we assume $P_{Net}(y)$ is symmetric in the x-z plane (see Fig.3) for the PKN ($a$ represents fracture half-height) and KGD ($a$ represents fracture half-length) fracture geometry, so that the odd function $G(\gamma)=0$. Because the fracture is discretized, the integral of Eq.(1), (3) and (4) have to be divided into different regions and then summed up. For example, the function of F($\gamma$) in the $m^{th}$ segment (see Fig.5) is calculated as:

$$F(\gamma)_m = -\frac{\gamma}{2\pi}(\int_{a_1}^{a_2}\frac{-P_{Net1}}{\sqrt{\gamma^2-\beta^2}}d\beta + \int_{a_2}^{a_3}\frac{-P_{Net2}}{\sqrt{\gamma^2-\beta^2}}d\beta + \cdots + \int_{a_m}^{\gamma}\frac{-P_{Netm}}{\sqrt{\gamma^2-\beta^2}}d\beta), \quad a_m \le y \le a_{m+1} \quad (A1)$$

Similarly, the integration to obtain $w_f(y)$ also needs to be integrated piecewise. The function of $w_f(y)$ in the $m^{th}$ segment is calculated as:

$$w_f(y)_m = \frac{16}{E'}(\int_y^{a_{m+1}}\frac{F(\gamma)_m}{\sqrt{\gamma^2-y^2}}d\gamma + \int_{a_{m+1}}^{a_{m+2}}\frac{F(\gamma)_{m+1}}{\sqrt{\gamma^2-y^2}}d\gamma + \cdots + \int_{a_n}^{a_{n+1}}\frac{F(\gamma)_n}{\sqrt{\gamma^2-y^2}}d\gamma), \quad a_m \le y \le a_{m+1} \quad (A2)$$

Substitute Eq.(A1) into Eq.(A2):

$$w_f(y)_m = \frac{8}{\pi E'}\int_y^{a_{m+1}}\left(\int_{a_1}^{a_2}\frac{\gamma P_{Net1}}{\sqrt{\gamma^2-\beta^2}\sqrt{\gamma^2-y^2}}d\beta + \int_{a_2}^{a_3}\frac{\gamma P_{Net2}}{\sqrt{\gamma^2-\beta^2}\sqrt{\gamma^2-y^2}}d\beta + \cdots + \int_{a_m}^{\gamma}\frac{\gamma P_{Netm}}{\sqrt{\gamma^2-\beta^2}\sqrt{\gamma^2-y^2}}d\beta\right)d\gamma$$

$$+ \int_{a_{m+1}}^{a_{m+2}}\left(\int_{a_1}^{a_2}\frac{\gamma P_{Net1}}{\sqrt{\gamma^2-\beta^2}\sqrt{\gamma^2-y^2}}d\beta + \int_{a_2}^{a_3}\frac{\gamma P_{Net2}}{\sqrt{\gamma^2-\beta^2}\sqrt{\gamma^2-y^2}}d\beta + \cdots + \int_{a_{m+1}}^{\gamma}\frac{\gamma P_{Net(m+1)}}{\sqrt{\gamma^2-\beta^2}\sqrt{\gamma^2-y^2}}d\beta\right)d\gamma + \cdots$$

$$+ \int_{a_n}^{a_{n+1}}\left(\int_{a_1}^{a_2}\frac{\gamma P_{Net1}}{\sqrt{\gamma^2-\beta^2}\sqrt{\gamma^2-y^2}}d\beta + \int_{a_2}^{a_3}\frac{\gamma P_{Net2}}{\sqrt{\gamma^2-\beta^2}\sqrt{\gamma^2-y^2}}d\beta + \cdots\right.$$

$$\left.+ \int_{a_n}^{\gamma}\frac{\gamma P_{Netn}}{\sqrt{\gamma^2-\beta^2}\sqrt{\gamma^2-y^2}}d\beta\right)d\gamma \quad (A3)$$

From Eq.(A3), we can observe that in the $m^{th}$ segment, fracture width, $w_f(y)_m$, is determined by the summation of numerous integrals, where the integration bounds and net pressure/stress $P_{Net}$ are different in each individual term. Because of the convoluted nature of the problem, it is difficult to observe the trend and establish a general algorithm to solve Eq.(A3). In addition, for the numerical integration (Shampine 2008) to work, Eq.(A3) has to be re-arranged, such that each single term is a definite double integral. To resolve this issue, we define the integration operator I:

$$I(L_\gamma, U_\gamma, L_\beta, U_\beta, P_{Net}) := \frac{8}{\pi E'}\int_{L_\gamma}^{U_\gamma}\int_{L_\beta}^{U_\beta}\frac{\gamma P_{Net}}{\sqrt{\gamma^2-\beta^2}\sqrt{\gamma^2-y^2}}d\beta d\gamma \quad (A4)$$

Re-write Eq.(A3) using the definition from Eq.(A4), we have

$$w_f(y)_m = [I(y, a_{m+1}, a_1, a_2, P_{Net1}) + I(y, a_{m+1}, a_2, a_3, P_{Net2}) + \cdots + I(y, a_{m+1}, a_m, \gamma, P_{Netm})] +$$

$$[I(a_{m+1}, a_{m+2}, a_1, a_2, P_{Net1}) + I(a_{m+1}, a_{m+2}, a_2, a_3, P_{Net2}) + \cdots + I(a_{m+1}, a_{m+2}, a_{m+1}, \gamma, P_{Net(m+1)})] +$$

$$\cdots + [I(a_n, a_{n+1}, a_1, a_2, P_{Net1}) + I(a_n, a_{n+1}, a_2, a_3, P_{Net2}) + \cdots + I(a_n, a_{n+1}, a_n, \gamma, P_{Netn})] \quad (A5)$$

Even though Eq.(A5) is a more compact formula, it is still difficult to observe the patterns and generalize a global algorithm to calculated fracture width with arbitrary values of $n$ and $P_{Net}$. Notice that $w_f(y)_m$ is the summation of the integration operator I that operated on different bounds and integrands, we put all the terms in the R.H.S of Eq.(A5) in to a matrix, called MatrixI(k,j), with k rows and j columns. Then Eq.(A5) becomes

$$w_f(y)_m = \sum MatrixI(k,j) \quad (A6)$$

Re-arrange MatrixI, such that

$$w_f(y)_m$$
$$= \sum \begin{bmatrix} I(y, a_{m+1}, a_1, a_2, P_{Net1}) & I(y, a_{m+1}, a_2, a_3, P_{Net2}) & \cdots & I(y, a_{m+1}, a_m, \gamma, P_{Netm}) \\ I(a_{m+1}, a_{m+2}, a_1, a_2, P_{Net1}) & I(a_{m+1}, a_{m+2}, a_2, a_3, P_{Net2}) & \cdots & I(a_{m+1}, a_{m+2}, a_{m+1}, \gamma, P_{Net(m+1)}) \\ \vdots & \vdots & \vdots & \vdots \\ I(a_n, a_{n+1}, a_1, a_2, P_{Net1}) & I(a_n, a_{n+1}, a_2, a_3, P_{Net2}) & \cdots & I(a_n, a_{n+1}, a_n, \gamma, P_{Netn}) \end{bmatrix} \quad (A7)$$

A closer observation of Eq.(A7), reveals that starting from the 2$^{nd}$ row, the outer integration bounds of operator I shift progressively from the $m^{th}$ segment to the $n^{th}$ segment; Excluding the last column, the inner integration bounds and $P_{Net}$ of operator I shift progressively from the 1$^{st}$ segment and the $(n-1)^{th}$ segment; In the first row, the lower bound of outer integration of operator I always starts from y, and in the last column, the upper bound of inner integration of operator I always ends with $\gamma$. Based on this analysis, an algorithm can be developed to calculated fracture width within the $m^{th}$ segment of fracture, and this process has to be repeated from the 1$^{st}$ segment to the $n^{th}$ segment to obtain the entire fracture width profile.

**Appendix-B: Determine Fracture Width with Arbitrary Normal Load for Radial Fracture Geometry**

Because the fracture is discretized in the radial direction, the integral of Eq.(11) has to be divided into different regions and then add up. For example, if $r$ belongs to the $m^{th}$ segment where $a_m \leq r \leq a_{m+1}$(see Fig.7), then

$$w_f(r_D)_m = \frac{8R_f}{\pi E'} \left\{ \left[ \int_{\frac{a_1}{R}}^{\frac{a_2}{R}} \frac{sP_{Net1}}{r_D} F\left(\sin^{-1}\sqrt{\frac{1-r_D^2}{1-s^2}}, \frac{s}{r_D}\right) ds + \int_{\frac{a_2}{R}}^{\frac{a_3}{R}} \frac{sP_{Net2}}{r_D} F\left(\sin^{-1}\sqrt{\frac{1-r_D^2}{1-s^2}}, \frac{s}{r_D}\right) ds + \cdots + \right.\right.$$

$$\left. \int_{\frac{a_m}{R}}^{r_D} \frac{sP_{Netm}}{r_D} F\left(\sin^{-1}\sqrt{\frac{1-r_D^2}{1-s^2}}, \frac{s}{r_D}\right) ds \right] + \left[ \int_{r_D}^{\frac{a_{m+1}}{R}} P_{Netm} F\left(\sin^{-1}\sqrt{\frac{1-s^2}{1-r_D^2}}, \frac{r_D}{s}\right) ds + \right.$$

$$\left.\left. \int_{\frac{a_{m+1}}{R}}^{\frac{a_{m+2}}{R}} P_{Net(m+1)} F\left(\sin^{-1}\sqrt{\frac{1-s^2}{1-r_D^2}}, \frac{r_D}{s}\right) ds + \cdots + \int_{\frac{a_n}{R}}^{\frac{a_{n+1}}{R}} P_{Netn} F\left(\sin^{-1}\sqrt{\frac{1-s^2}{1-r_D^2}}, \frac{r_D}{s}\right) ds \right] \right\} \quad (B1)$$

We define integration operators R1 and R2:

$$R1(L_s, U_s, P_{Net}) := \frac{8R_f}{\pi E'} \int_{L_s}^{L_s} \frac{sP_{Net}}{r_D} F\left(\sin^{-1}\sqrt{\frac{1-r_D^2}{1-s^2}}, \frac{s}{r_D}\right) ds \quad (B2)$$

$$R2(L_s, U_s, P_{Net}) := \frac{8R_f}{\pi E'} \int_{L_s}^{U_s} P_{Net} F\left(\sin^{-1}\sqrt{\frac{1-s^2}{1-r_D^2}}, \frac{r_D}{s}\right) ds \quad (B3)$$

We can rewrite Eq.(B1) using Eq.(B2) and (B3),to get:

$$w_f(r_D) = \left[ R1\left(\frac{a_1}{R_f}, \frac{a_2}{R_f}, P_{Net1}\right) + R1\left(\frac{a_2}{R_f}, \frac{a_3}{R_f}, P_{Net2}\right) + \cdots + R1\left(\frac{a_2}{R_f}, r_D, P_{Netm}\right) \right]$$

$$+ \left[ R2\left(r_D, \frac{a_{m+1}}{R_f}, P_{Netm}\right) + R2\left(\frac{a_{m+1}}{R_f}, \frac{a_{m+2}}{R_f}, P_{Net(m+1)}\right) + \cdots R2\left(\frac{a_n}{R_f}, \frac{a_{n+1}}{R_f}, P_{Netn}\right) \right] \quad (B4)$$

From Eq.(B4), we can observe that $w_f(r_D)$ can be expressed by the summation of two 1D matrices:

$$w_f(r_D)_m = \sum \text{MatrixR1} + \sum \text{MatrixR2} \quad (B5)$$

where

$$\text{MatrixR1} = \left[ R1\left(\frac{a_1}{R_f}, \frac{a_2}{R_f}, P_{Net1}\right) \ R1\left(\frac{a_2}{R_f}, \frac{a_3}{R_f}, P_{Net2}\right) \ \cdots \ R1\left(\frac{a_2}{R_f}, r_D, P_{Netm}\right) \right] \quad (B6)$$

$$\text{MatrixR2} = \left[ R2\left(r_D, \frac{a_{m+1}}{R_f}, P_{Netm}\right) \ R2\left(\frac{a_{m+1}}{R_f}, \frac{a_{m+2}}{R_f}, P_{Net(m+1)}\right) \ \cdots \ R2\left(\frac{a_n}{R_f}, \frac{a_{n+1}}{R_f}, P_{Netn}\right) \right] \quad (B7)$$

By examining Eq.(B5) to Eq.(B7), an algorithm can be developed to calculated fracture width within the $m^{th}$ segment of the fracture, and this process needs to be repeated from the $1^{st}$ segment to the $n^{th}$ segment to obtain the entire fracture width profile.